\documentclass[eqsecnum,preprint,pre,aps,showpacs]{revtex4}
\usepackage{graphicx}
\usepackage{amsmath}
\begin{document}

\title{Simplified Self-Consistent Theory of Colloid Dynamics}
\author{R. Ju\'arez-Maldonado$^{1}$, M. A. Ch\'avez-Rojo$^{2}$,
P. E. Ram\'{\i}rez-Gonz\'alez$^{1}$, L. Yeomans-Reyna$^{3}$,
and M. Medina-Noyola$^1$}
\address{\\$^{1}$Instituto de F\'{\i}sica {\sl ``Manuel Sandoval Vallarta"},
Universidad Aut\'{o}noma de San Luis Potos\'{\i}, \'{A}lvaro
Obreg\'{o}n 64, 78000 San Luis Potos\'{\i}, SLP, M\'{e}xico\\
$^{2}$Facultad de Ciencias Qu\'{\i}micas, Universidad Aut\'onoma de
Chihuahua, Venustiano Carranza S/N, 31000 Chihuahua, Chih.,
M\'exico.\\
$^{3}$Departamento de F\'{\i}sica, Universidad de Sonora, Boulevard Luis\\
Encinas y Rosales, 83000, Hermosillo, Sonora, M\'{e}xico.}

\date{\today}

\begin{abstract}
One of the main elements of the self-consistent generalized Langevin
equation (SCGLE) theory of colloid dynamics [Phys. Rev. E {\bf 62},
3382 (2000); ibid {\bf 72}, 031107 (2005)] is the introduction of
exact short-time moment conditions in its formulation. The need to
previously calculate these exact short-time properties constitutes a
practical barrier for its application. In this note we report that a
simplified version of this theory, in which this short-time
information is eliminated, leads to the same results in the
intermediate and long-time regimes. Deviations are only observed at
short times, and are not qualitatively or quantitatively important.
This is illustrated by comparing the two versions of the theory
for representative model systems.
\end{abstract}

\pacs{64.70.Pf, 61.20.Gy, 47.57.J-}

\maketitle

\section{Introduction}

In recent work a new first-principles theory of dynamic arrest has
been proposed \cite{rmf, todos1}. This consists essentially of the
application of the self-consistent generalized Langevin equation
(SCGLE) theory of colloid dynamics \cite{5,4,3,TAVSDB} to the
description of the singular behavior characteristic of dynamic
arrest phenomena in specific colloidal systems and conditions. The
SCGLE theory was originally devised to describe tracer and
collective diffusion properties of colloidal dispersions in the
short- and intermediate-times regimes \cite{1,6}. Its
self-consistent character, however, introduces a non-linear dynamic
feedback, leading to the prediction of dynamic arrest in these
systems, similar to that exhibited by the mode coupling theory (MCT)
of the ideal glass transition \cite{goetze1}. The resulting theory
of dynamic arrest in colloidal dispersions was applied in recent
work to describe the glass transition in three mono-disperse
experimental model colloidal systems with specific (hard-sphere,
screened electrostatic, and depletion) inter-particle effective
forces \cite{rmf, todos1}. The results indicate that the SCGLE
theory of dynamic arrest has the same or better level of
quantitative predictive power as conventional MCT, but is built on a
completely independent conceptual basis, thus providing an
alternative approach to the description of dynamic arrest phenomena.

There is, however, a possible practical disadvantage of the SCGLE
with respect to the MCT, and it refers to the fact that the MCT only
requires the static structure factor of the system as an external
input, whereas the SCGLE theory requires this information plus other
additional static properties involved in the exact short-time
conditions that the theory has built-in \cite{3}. As it happens,
however, the long-time asymptotic solutions of the relaxation
equations that constitutes the SCGLE theory are independent of such
exact short-time properties \cite{todos1}. The questions then arise
if a simplified version of the SCGLE theory, in which this
short-time information is eliminated, could be proposed, and to what
extent such a simpler theory will still provide a reliable
representation of the dynamics of the colloidal system not only in
the asymptotic long-time regime, but also at earlier stages. In what
follows we demonstrate that there is a simple manner to build this
simplified version of the SCGLE theory, and that it is virtually as
accurate as the full version, even in the short- and
intermediate-time regimes. This finding will greatly simplify the
application of the SCGLE theory of dynamic arrest.

Let us summarize the four distinct fundamental elements of the full
self-consistent generalized Langevin equation theory of colloid
dynamics. The first consists of general  and exact memory-function
expressions for the intermediate scattering function $F(k,t)$ and
its self component $F_{S}(k,t)$, derived with the generalized
Langevin equation (GLE) formalism \cite{16}, which in Laplace space
read \cite{5}

\begin{equation}
F(k,z)=\frac{S(k)}{z+\frac{k^{2}D_{0}S^{-1}(k)}{1+C(k,z)}},
\label{fkz}
\end{equation}

\begin{equation}
F_{S}(k,z)=\frac{1}{z+\frac{k^{2}D_{0}}{1+C_{S}(k,z)}}, \label{fskz}
\end{equation}

\noindent where $D_0$ is the free-diffusion coefficient, $S(k)$ is
the static structure factor of the system, and $C(k,z)$ and
$C_{S}(k,z)$ are the corresponding memory functions.

The second element is an approximate relationship between collective
and self-dynamics. In the original proposal of the SCGLE theory
\cite{5}, two possibilities, referred to as the additive and the
multiplicative Vineyard-like approximations, were considered. The
first approximates the difference $[C(k,t)-C_{S}(k,t)]$, and the
second the ratio $[C(k,t)/C_{S}(k,t)]$, of the memory functions, by
their exact short-time limits, using the fact that the exact
short-time expressions for these memory functions, denoted by
$C^{SEXP}(k,t)$ and $C_{S}^{SEXP}(k,t)$, are known in terms of
equilibrium structural properties \cite{3,21}. The multiplicative
approximation was devised to describe more accurately the very early
relaxation of $F(k,t)$ \cite{TAVSDB}, but the additive approximation
was found to provide a more accurate prediction of dynamic arrest
phenomena \cite{todos1}. In this paper, for ``full SCGLE theory" we
refer to the theory that involves the \textit{additive}
Vineyard-like approximation,

\begin{equation}
C(k,t)  = C_{S}(k,t) + [C^{SEXP}(k,t)- C_{S}^{SEXP}(k,t)].
\label{additive}
\end{equation}

The third ingredient consists of the independent approximate
determination of $F_{S}(k,t)$ [or $C_{S}(k,t)$]. One intuitively
expects that these $k$-dependent self-diffusion properties should be
simply related to the properties that describe the Brownian motion
of individual particles, just like in the Gaussian approximation
\cite{1}, which expresses $F_{S}(k,t)$ in terms of the mean-squared
displacement (msd) $ \overline{(\Delta x(t))^2}$ as $
F_{S}(k,t)=\exp[-k^{2} \overline{(\Delta x(t))^2}/2]$. We introduce
an analogous approximate connection, but at the level of their
respective memory functions. The memory function of $
\overline{(\Delta x(t))^2}$ is the so-called time-dependent friction
function $\Delta \zeta (t)$. This function, normalized by the
solvent friction $\zeta_0$, is the \emph{exact} long wave-length
limit of $C_{S}(k,t)$, i.e., $\lim_{k\to0} C_{S}(k,t)= \Delta
\zeta^* (t)\equiv \Delta \zeta(t)/\zeta_0$. Thus, we interpolate
$C_{S}(k,t)$ between its two limits, namely,

\begin{equation}
C_{S}(k,t)=C_{S}^{SEXP}(k,t)+\left[ \Delta \zeta^*
(t)-C_{S}^{SEXP}(k,t)\right] \lambda (k), \label{interpolation}
\end{equation}

\noindent where

\begin{equation}
\lambda (k)\equiv [1+(k/k_c)^2]^{-1} \label{lambdadk}
\end{equation}
is a phenomenological interpolating function, with $k_c$ being the
position of the first minimum that follows the main peak of $S(k)$
\cite{3}.

The fourth ingredient of our theory is another exact result, also
derived within the GLE approach \cite{16}, this time for $\Delta
\zeta^* (t)$. This exact result may, upon a well-defined simplifying
approximation, be converted into the following approximate but
general expression \cite{todos1}

\begin{equation}
\Delta \zeta^* (t) =\frac{D_0}{3\left( 2\pi \right) ^{3}n}\int d
{\bf k}\left[\frac{ k[S(k)-1]}{S(k)}\right] ^{2}F(k,t)F_{S}(k,t).
\label{dzdt}
\end{equation}

Eqs.\ (\ref{fkz})--(\ref{dzdt})  constitute the full SCGLE theory of
colloid dynamics. Besides the unknown dynamic properties, it
involves the equilibrium properties $S(k)$, $C^{SEXP}(k,t)$ and
$C_{S}^{SEXP}(k,t)$, determined by the methods of equilibrium
statistical thermodynamics. We should also point out that
Eqs.\ (\ref{fkz}) and (\ref{fskz}) are exact results, and that
Eq.\ (\ref{dzdt}) derives from another exact result. Hence, it should not
be a surprise that the same results are used by other theories; in
fact, the same equations are employed in MCT. The difference lies,
of course, in the the manner we relate and use them. In this sense,
the distinctive elements of the SCGLE theory are the Vineyard-like
approximation in Eq.\ (\ref{additive}) and the interpolating
approximation in Eq.\ (\ref{interpolation}).

The simplified version of the SCGLE theory is now suggested by the
form that these distinctive equations (Eqs.\ (\ref{additive}) and
(\ref{interpolation})) attain for times longer than the relaxation
time of the functions $C^{SEXP}(k,t)$ and $C_{S}^{SEXP}(k,t)$. Under
those conditions, Eqs.\ (\ref{additive}) and (\ref{interpolation})
become, respectively,

\begin{equation}
C(k,t)  = C_{S}(k,t). \label{simpleradditive}
\end{equation}
and

\begin{equation}
C_{S}(k,t)=\left[ \Delta \zeta^* (t)\right] \lambda (k).
\label{simplerinterpolation}
\end{equation}

It is not difficult to see that the original self-consistent set of
equations (involving Eqs.\ (\ref{additive}) and
(\ref{interpolation})) shares the same long-time asymptotic
stationary solutions as its simplified version. Such stationary
solutions are given by \cite{todos1}

\begin{equation}
\lim_{t\to\infty}F(k,t)  = \frac
{\lambda(k)S(k)}{\lambda(k)S(k)+k^2\gamma}S(k). \label{fdkinf}
\end{equation}
and

\begin{equation}
\lim_{t\to\infty}F_S(k,t)  = \frac
{\lambda(k)}{\lambda(k)+k^2\gamma}. \label{fdksinf}
\end{equation}
where $\gamma$ is the solution of the following equation

\begin{equation}
\frac{1}{\gamma} = \frac{1}{6\pi^{2}n}\int_{0}^{\infty }
dkk^4\frac{\left[S(k)-1\right] ^{2}\lambda^2 (k)}{\left[\lambda
(k)S(k) + k^2\gamma\right]\left[\lambda (k) + k^2\gamma\right]}.
\label{nep5pp}
\end{equation}
The parameter $\gamma$ is the long-time asymptotic value, of the msd, i.e.,
$\gamma \equiv \lim_{t \to \infty} \overline{(\Delta x(t))^2}$.
In the arrested states, this parameter is finite, representing the
localization of the particles, whereas in the ergodic states it
diverges.

It is then natural to ask what the consequences would be of
replacing Eqs.\ (\ref{additive}) and (\ref{interpolation}) of the
full SCGLE set of equations by the simpler approximations in
Eqs.\ (\ref{simpleradditive}) and (\ref{simplerinterpolation}), that no
longer contain the functions $C^{SEXP}(k,t)$ and
$C_{S}^{SEXP}(k,t)$. Our proposal of a simplified version of the
SCGLE theory consists precisely of this replacement, so that the
``simplified SCGLE theory" consists of the exact results in
Eqs.\ (\ref{fkz}) and (\ref{fskz}) along with Eqs.\ (\ref{lambdadk}) and
(\ref{dzdt}), complemented by the closure approximations in
Eqs.\ (\ref{simpleradditive}) and (\ref{simplerinterpolation}).

We have made a systematic comparison of the various dynamic
properties involved in the SCGLE theory, including the intermediate
scattering function $F(k,t)$, its self component $F_{S}(k,t)$, and
other tracer-diffusion properties such as the time-dependent
friction function $\Delta \zeta^* (t)$, the mean squared
displacement or the time-dependent diffusion coefficient $D(t)\equiv
 \overline{(\Delta x(t))^2}/2t$. As expected, the scenario of dynamic
arrest exhibited by this simpler theory is identical to that
provided by the full SCGLEscheme. This is probably not surprising
since, as indicated above, both sets of dynamic equations share the
same long-time asymptotic behavior and the same asymptotic
stationary solutions. What is surprising, however, is the degree of
accuracy of the simplified theory in the short- and
intermediate-time regimes. In our systematic comparison we
considered systems with soft-sphere, hard-sphere, and repulsive
Yukawa interactions, systems with attractive (Yukawa) interactions,
systems in three and in two dimensions, and both, mono-disperse and
bi-disperse systems, in all cases with similar conclusions, that we
illustrate with the following examples.

\begin{figure}[ht]
\includegraphics[scale=.25]{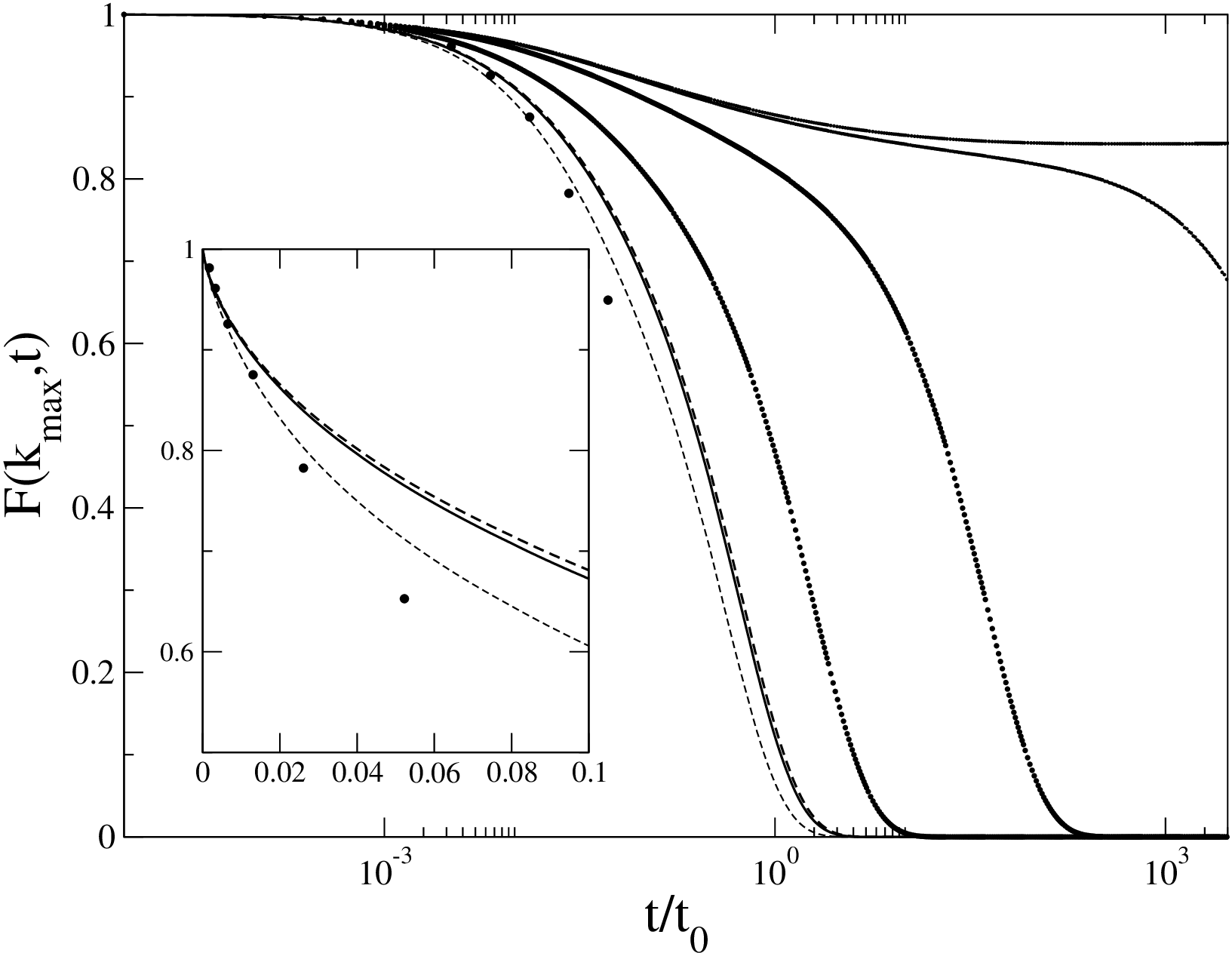}
\caption{$F(k,t)$ as a function of time (in units of $t_0\equiv
\sigma^2/D_0$) at the position $k_{max}$ of the first maximum of
$S(k)$ of a soft-sphere system with $\nu=18$ and volume fractions
$\phi=0.515, 0.56, 0.60, 0.612,$ and $0.613$, calculated with the
simplified (heavy solid line) and the full (heavy dashed line)
versions of the SCGLE theory. For $\phi=0.515$ we also show the
intermediate-time Brownian dynamics data (solid circles) and the
results of the SCGLE theory with the multiplicative approximation
reported in Ref.\ \cite{TAVSDB}. The inset is a close up view in
linear scale of the short-time relaxation of $F(k,t)$. }
\label{fig.1}
\end{figure}

Thus, in Fig.\ \ref{fig.1} we plot the intermediate scattering function
$F(k,t)$ as a function of time, evaluated at the position $k_{max}$
of the first maximum of the static structure factor of a soft-sphere
system. The pair potential, in units of the thermal energy
$k_{B}T=\beta ^{-1}$, is given by $ \beta u(r)=\frac{1}{(r/\sigma
)^{2\nu }}-\frac{2}{(r/\sigma )^{\nu }}+1$ for $0<r<\sigma$, and it
vanishes for $r>\sigma$. The system in Fig.\ \ref{fig.1} corresponds to
$\nu=18$ and to the volume fractions $\phi\equiv \pi n \sigma^3/6
=0.515$, $ 0.612$, and $0.613$, and the static structure factor was
calculated using the prescription of Verlet and Weis
\cite{verletweis}. The heavy-solid lines correspond to the
simplified version, and the heavy-dashed lines to the full version,
of the SCGLE theory. The very first feature to notice is the virtual
coincidence of the results of these two approximations; in fact,
only for $\phi=0.515$ the difference is appreciable. For the other
volume fractions the results are not distinguishable in the scale of
the figure, and this includes the vicinity of the glass transition
which, as can be seen in the figure, is predicted to occur at
$\phi_g=0.613$ for this soft-sphere system. For $\phi=0.515$ we also
show the Brownian dynamics data (solid circles) reported in
Ref.\ \cite{TAVSDB}, as well as the results of the SCGLE theory within the
\textit{multiplicative} Vineyard-like approximation (soft dashed
lines), to recall the fact that the multiplicative approximation
sometimes provides a slightly more accurate quantitative description
of the initial relaxation of $F(k,t)$ (see the inset). In a longer
time-scale, as indicated in the main figure, the prediction of the
overall relaxation provided by the SCGLE theory complemented with
the additive and the multiplicative approximations is quite similar.
Furthermore, as discussed in Ref.\ \cite{todos1}, the additive
approximation provides a simpler and more accurate description of
dynamic arrest, partly because these phenomena do not seem to depend
strongly on the short-time behavior illustrated in the inset of
Fig.\ \ref{fig.1}. Thus, from now on, we shall omit further
reference to the multiplicative approximation.

\begin{figure}[ht]
\includegraphics[scale=.3]{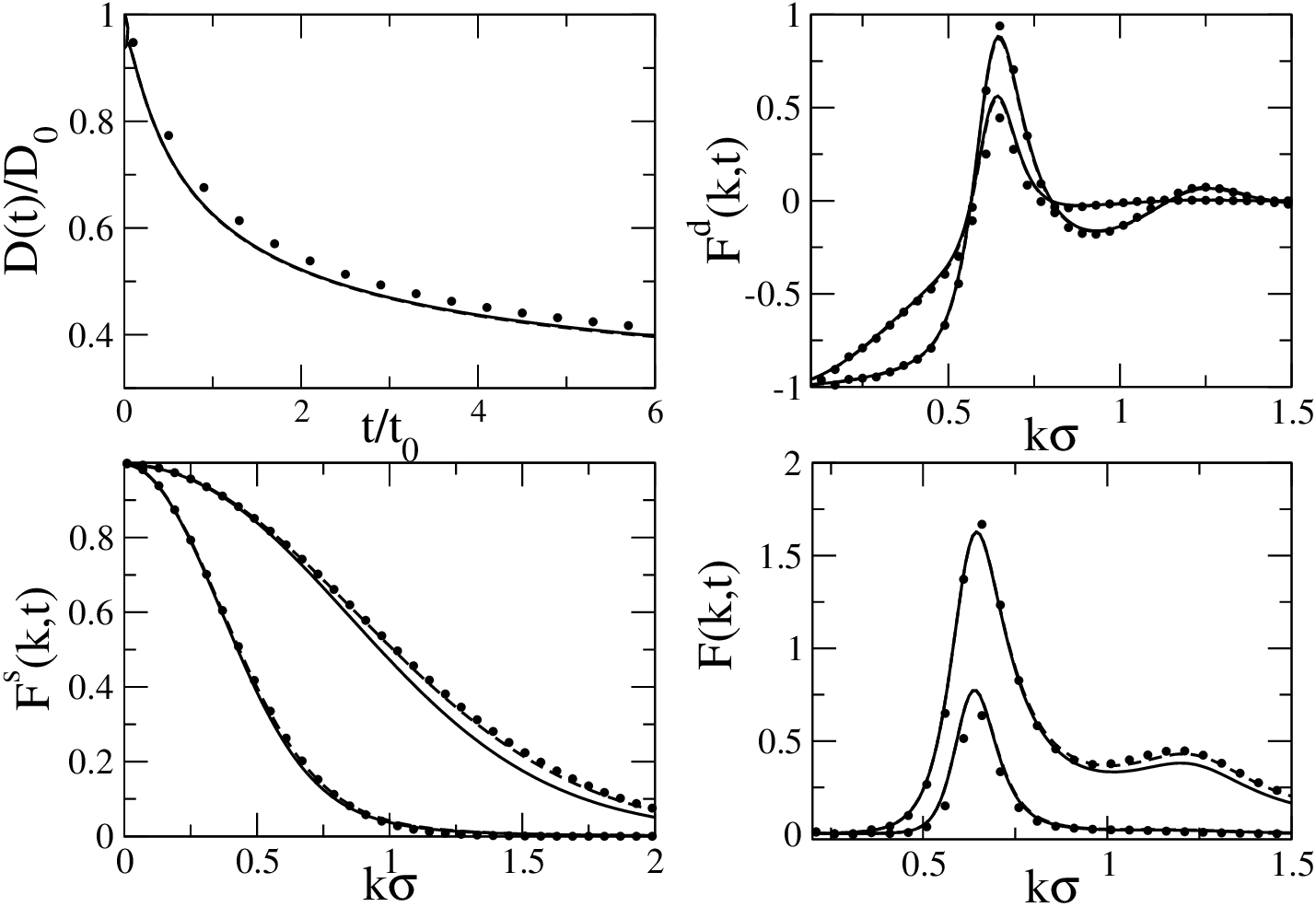}
\caption{Time-dependent diffusion coefficient $D(t)$  and (self,
distinct, and total) intermediate scattering functions $F^s(k,t)$,
$F^d(k,t)$ and $F(k,t)$ of a repulsive Yukawa system with $z=0.15$,
$K=500$, and $\phi=4.4\times10^{-4}$. Results of the simplified (solid lines)
and of the full (dashed lines) SCGLE theory for the intermediate
scattering functions are shown for times $t=t_0$ (upper or more
structured curves) and $t=10t_0$. The solid circles are the Brownian
dynamic data of Ref.\ \cite{marco1}. } \label{fig.2}
\end{figure}

A similar situation is illustrated in Fig.\ \ref{fig.2},
this time for a system
of colloidal particles interacting though a hard-sphere potential of
diameter $\sigma$ plus an additional long-ranged repulsive Yukawa
tail of the form $\beta u(r) = K \exp{-z(r/\sigma -1)}/(r/\sigma)$,
with $z=0.15$ and $K=500$. For a volume fraction $\phi_1=4.4\times10^{-4}$,
it corresponds to the conditions of Fig.\ 1 of Ref.\ \cite{marco1}. In
our present figure, however, we compare the full and the simplified
SCGLE theory for the time-dependent diffusion coefficient $D(t)$ and
for the intermediate scattering functions $F(k,t)$, $F^s(k,t)$, and
$F^d(k,t)(\equiv F(k,t)-F^s(k,t))$ in the short-time regime, where
these differences are expected to be larger. Our present comparison
indicates that the new simplified version of the SCGLE theory leads
to essentially identical results, even in these time regimes.

\begin{figure}[ht]
\includegraphics[scale=.3]{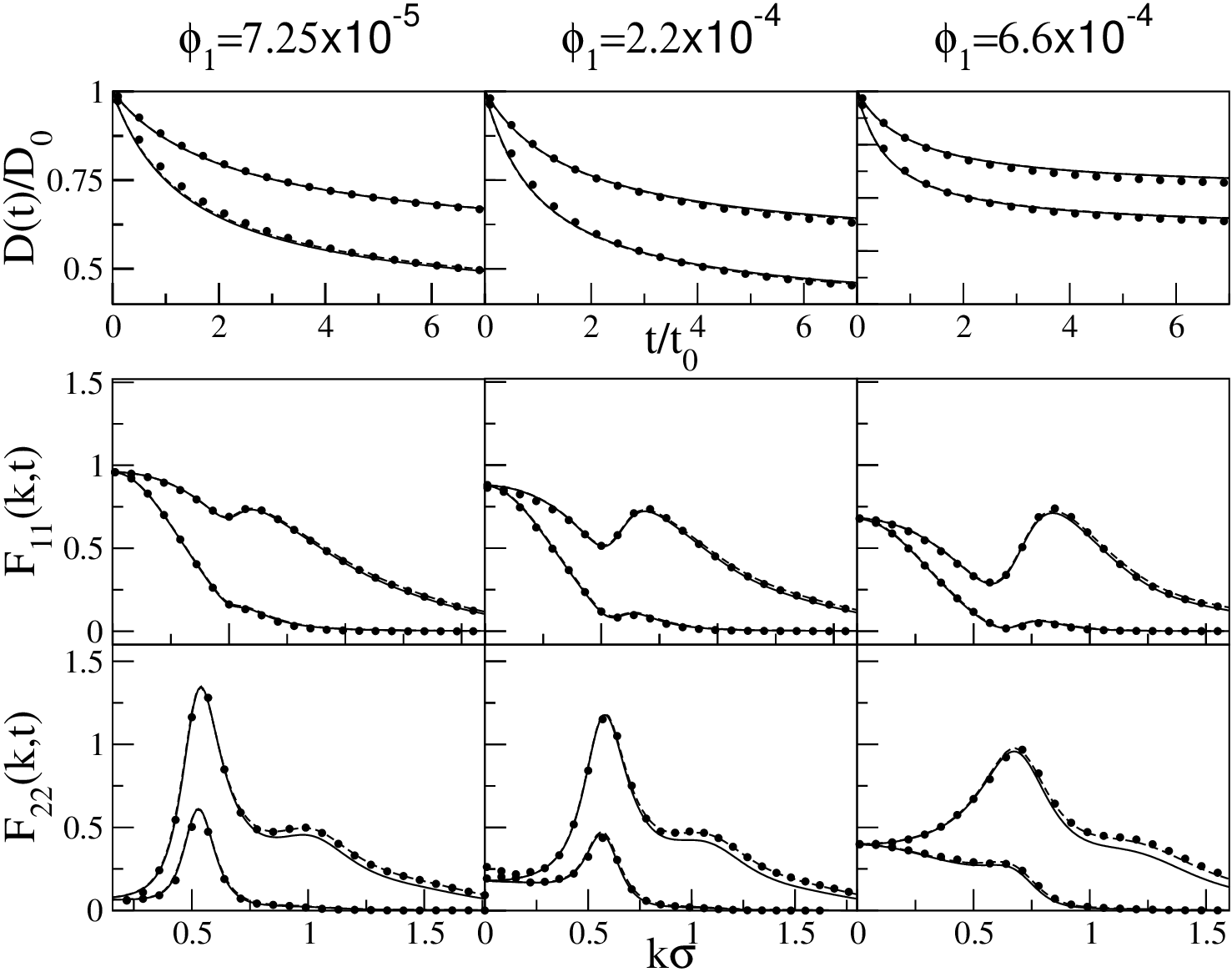}
\caption{ Time-dependent diffusion coefficients $D_1(t)$ and
$D_2(t)$ as a function of time (upper curve corresponding to species
1), and total intermediate scattering functions $F_{11}(k,t)$ and
$F_{22}(k,t)$ of a repulsive Yukawa mixture with $z=0.15$,
$K_1=100$, $K_2=500$ for $t=t_0$ and $t=10t_0$. The volume fraction
of the more interacting species is kept fixed at $\phi_2=2.2\times10^{-4}$
and $\phi_1$ takes the values $\phi_1=7.25\times10^{-5}$ (left column),
$2.2\times10^{-4}$(center column), and $6.6\times10^{-4}$(rigth column).
The symbology and conventions are the same as in Fig.\ \ref{fig.2}.}
\label{fig.3}
\end{figure}

Just like the MCT has been extended to mixtures \cite{bossethakur1},
the SCGLE has also been extended to multi-component colloidal
systems \cite{marco1, marco2}. Also in this case the simplified
version of the SCGLE theory provides virtually the same description
as the full SCGLE scheme, but its practical application is far
simpler. In Fig.\ \ref{fig.3} a comparison is presented for a binary
Yukawa mixture, with pair potential $\beta u_{ij}(r) = \sqrt{K_iK_j}
\exp{-z(r/\sigma -1)}/(r/\sigma)\ \ (1\le i,j\le 2)$ in which a
fraction $x_1$ of the particles (species $1$) interact with a charge
parameter $K_1=100$, the other fraction $x_2$ (species 2) with
$K_2=500$. The volume fraction of the more interacting species is
kept fixed at $\phi_2=2.2\times10^{-4}$ and $\phi_1$ takes the
values $6.6\times10^{-4}$ (right column), $2.2\times10^{-4}$ (middle
column), and $7.25\times10^{-5}$(left column), corresponding to
$x_1=0.75$, $x_1=0.5$ and $x_1=0.25$. This figure corresponds to the
same conditions as Fig.\ 3 of Ref.\ \cite{marco2} and, as in our
previous example, the simulated $S(k)$ was employed. It simply
confirms the general conclusions of this communication, namely, that
the simplified SCGLE theory provides a description of the relaxation
of concentration fluctuations in colloidal suspensions qualitatively
and quantitatively virtually identical to the full SCGLE theory. Its
practical implementation, however, is much simpler than either the
full SCGLE or the MCT schemes. This has simplified the application
of the SCGLE theory to the discussion of dynamic arrest in colloidal
mixtures  \cite{mixtures1} and in colloidal fluids adsorbed in model
porous media \cite{porous1}, that we report in separate
communications.

\section*{ACKNOWLEDGMENTS:}
 This work was supported by the Consejo
Nacional de Ciencia y Tecnolog\'{\i}a (CONACYT, M\'{e}xico), through
grant No. 2004-C01-47611, 2003-C02-44744 and 2006-C01-60064, 
and by FAI-UASLP.

\newpage
\section*{FIGURES AND CAPTIONS:}

\end{document}